\documentclass[11pt]{article}
\usepackage{amsmath,amssymb}

\usepackage[english]{babel} 
\usepackage[T1]{fontenc}
\usepackage{amsmath, amsthm}
\usepackage[all,dvips]{xy}
\usepackage{booktabs} 

\usepackage{tikz}
\usepackage{tikz,fullpage}
\usetikzlibrary{matrix}
\usepackage{authblk}
\usepackage{geometry}
\usepackage{enumerate}
\usepackage{stmaryrd}
\usepackage{array}
\usepackage{graphicx}
\usepackage{color}
\definecolor{vert}{rgb}{0.02,0.4,0.10}
\usepackage{hyperref}
\hypersetup{
    colorlinks=true,
    breaklinks=true,
    linkcolor=vert,
    citecolor=blue
}

\usepackage{graphicx}
\graphicspath{ {images/} }
\usepackage{pgf,tikz,pgfplots}
\pgfplotsset{compat=1.14}
\usepackage{mathrsfs}
\usetikzlibrary{arrows}

\usepackage{makeidx}
\usepackage{pstricks}

\newcommand{\SN}[2]{[\! [{#1},{#2}]\! ]}
\newcommand{\N}{\mathbb{N}}
\newcommand{\R}{\mathbb{R}}

\newtheorem{prop}{Proposition}

\theoremstyle{remark}
\newtheorem{definition}{Definition}
\newtheorem{rem}{Remark}

\usepackage{enumitem}
\numberwithin{equation}{subsection}

\usepackage[backend=biber, style=alphabetic]{biblatex}
\usepackage{csquotes}
\title{Mesh Denoising}
\date{January 2022}
\author[1, 2, 3]{Constantin Vaillant-Tenzer}
\affil[1]{École Normale Supérieure - PSL: Department of Cognitive Studies}
\affil[2]{École Normale Supérieure - PSL: Department of Mathematics and Applications}
\affil[3]{Université Paris Cité: UFR Fundamental and Biomedical Sciences and IHSS}
\affil{constantin.tenzer@ens.psl.eu}

\begin{document}

\maketitle

\begin{abstract}
In this paper, we study four mesh denoising methods: linear filtering, a heat diffusion method, Sobolev regularization, and, to a lesser extent, a barycentric approach based on the Sinkhorn algorithm. We illustrate that, for a simple image denoising task, a naive choice of a Gibbs kernel can lead to unsatisfactory results. We demonstrate that while Sobolev regularization is the fastest method in our implementation, it produces slightly less faithful denoised meshes than the best results obtained with iterative filtering or heat diffusion. We empirically show that, for the large mesh considered, the heat diffusion method is slower and not more effective than filtering, whereas on a small mesh an appropriate choice of diffusion parameters can improve the quality. Finally, we observe that all three mesh-based methods perform markedly better on the large mesh than on the small one.
\end{abstract}

\section{Introduction}

\subsection{The Problem}

\begin{figure}[h]
\label{probfig}
      \centering
      \includegraphics[scale=0.2]{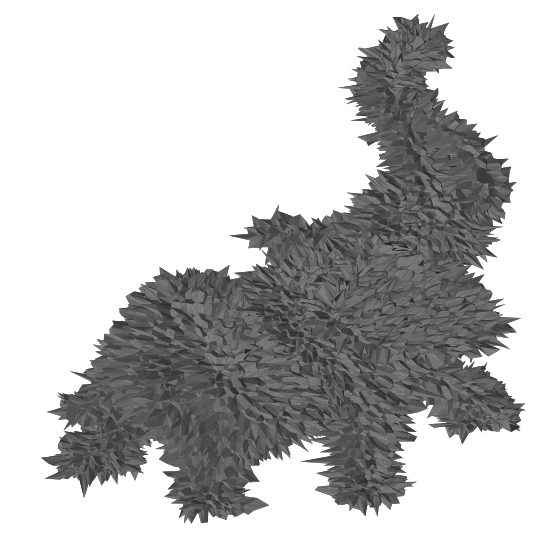}
      \includegraphics[scale=0.2]{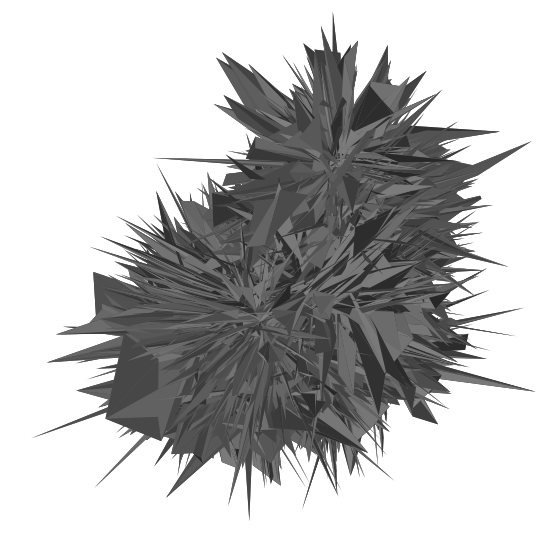} \\
           \includegraphics[scale=0.2]{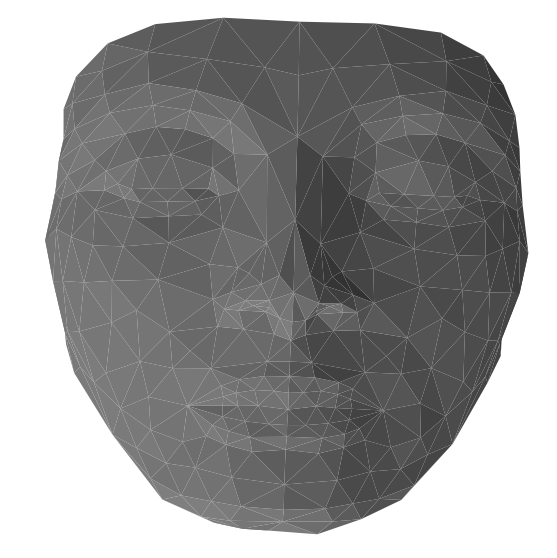}
       \includegraphics[scale=0.2]{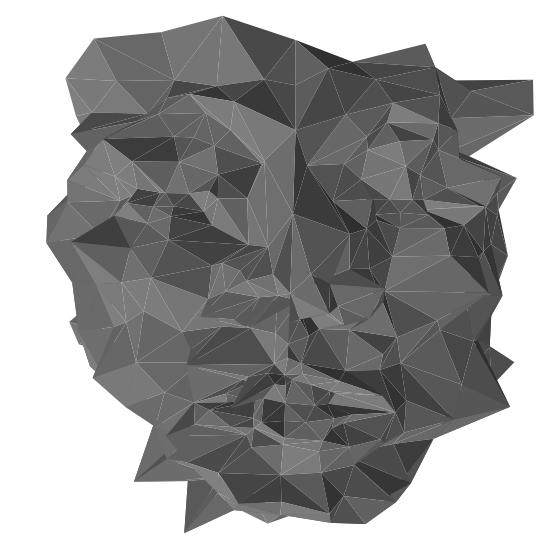}
            \includegraphics[scale=0.2]{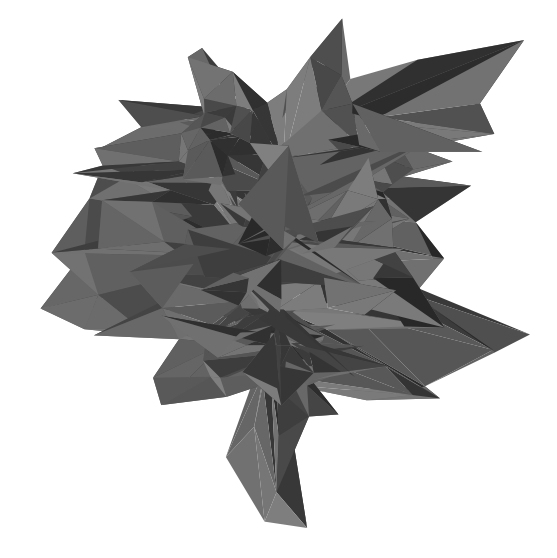}
      \caption{The various noisy meshes studied in Table \ref{datamaillage}. From left to right: $\rho = 0.015$, $\rho = 0.2$, $\rho = 1$ (bottom row only).}
\end{figure}

To represent three-dimensional objects within a computer, meshes are highly practical. A mesh $\mathcal{M}$ consists of an index set $\mathcal{V} = \SN{1}{n}$, which indexes $n$ vertices, a set of edges $\mathcal{E} \subset \mathcal{V} \times \mathcal{V}$, and a set of faces $\mathcal{F} \subset \mathcal{V} \times \mathcal{V} \times \mathcal{V}$. Although these objects are not yet ubiquitous in everyday applications, they are natural candidates for representing and transmitting 3D content, for instance in holographic displays or immersive communications. Since information transmitted through a noisy channel can be corrupted, denoising becomes necessary. For our study, we start with the noisy meshes shown in Figure~\ref{probfig}.

However, this task is not straightforward. First, these objects can have a large number of vertices $n$, especially when they aim to faithfully represent a three-dimensional real-world object. Thus, an algorithm must be sufficiently fast and scalable. It may also be desirable for a machine to determine whether a received mesh needs denoising and to estimate the noise level without knowing the original. These are learning problems that are not addressed here. We instead focus on algorithmic denoising methods that could serve as building blocks in supervised learning pipelines.

One promising but technically demanding approach relies on optimal transport on geometric domains, implemented numerically via the Sinkhorn algorithm~\cite{SOL2015}. Faster or simpler methods, such as Sobolev regularization or graph-based filtering, can nevertheless yield good results. It is important to understand their limitations and regimes of effectiveness. In particular, we will see that they behave quite differently on small and large meshes, yet can still turn severely degraded shapes into visually interpretable ones.

\subsection{Contributions}

After reviewing the theoretical foundations of the Sinkhorn algorithm---namely discrete optimal transport and the Kullback--Leibler divergence---and recalling the heat equation and Cholesky factorization, we briefly attempt to apply an entropic optimal transport scheme to denoise an image. We then focus on three mesh denoising methods: a low-pass graph filter, a linear heat diffusion method, and a Sobolev-type regularization. We compare their accuracy and execution time on two meshes of very different sizes ($n = 24{,}955$ and $n = 299$ vertices). Finally, we discuss the main limitations observed in our experiments and outline possible extensions.

\section{Theory Surrounding the Use of the Sinkhorn Algorithm}

\subsection{Optimal Transport}
\label{transopt}

\subsubsection{The Monge Problem and Its Relaxation by Kantorovich}

The Monge problem is a discrete optimization problem.

\begin{definition}
Let $X$ and $Y$ be two sets of $N \geq 1$ elements. We define a cost matrix $c := (c_{x,y})_{(x,y) \in X \times Y} \in \R^{X \times Y}$. We denote by $\mathrm{Bij}(X,Y)$ the set of bijections from $X$ to $Y$.

The Monge problem is expressed as:
\begin{equation}
\min_{\varphi \in \mathrm{Bij}(X,Y)} \sum_{x \in X} c_{x,\varphi(x)}.
\end{equation}
\end{definition}

\begin{rem}
The sets $X$ and $Y$ are often sets of points in the same vector space. Initially, there is an object at each point in $X$. For $(x,y) \in X \times Y$, $c_{x,y}$ models the cost of transporting the object from $x$ to $y$. The goal is to transport the objects from $X$ to $Y$, placing one object at each destination point, while minimizing the total cost. Given this interpretation, it is natural (though not strictly required for the basic existence theory) to assume that $c_{x,y} \geq 0$.
\end{rem}

\begin{rem}
Expressed in this way, the problem is purely discrete, and $\mathrm{Bij}(X,Y)$ is finite. This leads to a combinatorial optimization problem whose complexity grows rapidly with $N$. A classical way to facilitate its resolution is to consider a relaxation in terms of transport plans.
\end{rem}

The Monge--Kantorovich problem is, a priori, more general. Let $X$ and $Y$ be two non-empty finite sets with cardinalities $N$ and $M$, respectively. Let $\mu$ and $\nu$ be probability measures on $X$ and $Y$, respectively. The goal is to transport all the mass from $X$ to $Y$ while respecting the initial mass at each $x$ and the prescribed mass at each $y$.

\begin{definition}
A transport plan in this context is defined as follows: $\gamma := (\gamma_{x,y})_{(x,y) \in X \times Y}$, a family of non-negative real numbers, is a transport plan if $\gamma \in \Pi_{\mu,\nu}$, where:
\begin{equation}
\Pi_{\mu,\nu} := \Bigl\{ \gamma \in \R_+^{X \times Y} \,\Bigm|\, \forall x \in X,\; \sum_{y \in Y} \gamma_{x,y} = \mu_x,\;\; \forall y \in Y,\; \sum_{x \in X} \gamma_{x,y} = \nu_y \Bigr\}.
\end{equation}
\end{definition}

\begin{definition}
The cost functional for the Monge--Kantorovich problem, using the same cost family $c$ as for Monge, is defined as:
\begin{equation}
C(\gamma) := \sum_{(x,y) \in X \times Y} c_{x,y}\,\gamma_{x,y}.
\end{equation}
\end{definition}

\begin{definition}
The Monge--Kantorovich (MK) problem is expressed as:
\begin{equation}
\min_{\gamma \in \Pi_{\mu,\nu}} C(\gamma).
\end{equation}
\end{definition}

\begin{prop}
The (MK) problem admits a solution \cite{OTAM}.
\end{prop}

\begin{proof}
The set $\Pi_{\mu,\nu}$ is non-empty, for instance because the product measure $\mu \otimes \nu$ defines a feasible transport plan. The set $\Pi_{\mu,\nu}$ is closed and convex. Since $\mu$ and $\nu$ are probability measures, all entries satisfy $0 \leq \gamma_{x,y} \leq 1$, and the total mass is fixed, so $\Pi_{\mu,\nu}$ is bounded in $\R^{X\times Y}$. Thus, $\Pi_{\mu,\nu}$ is compact. The functional $C$ is linear in $\gamma$, hence continuous. The existence of a minimum follows from the Weierstrass theorem.
\end{proof}

\begin{rem}
Assume that $X$ and $Y$ have the same cardinality $N$ and that $\mu$ and $\nu$ are uniform on $X$ and $Y$ respectively. Identifying permutations with permutation matrices via
\[
\varphi \in \mathrm{Bij}(X,Y) \;\longmapsto\; (\mathbb{1}_{\{\varphi(x)=y\}})_{x,y},
\]
one can verify that
\begin{equation}
\arg\min_{\varphi \in \mathrm{Bij}(X,Y)} \sum_{x \in X} c_{x,\varphi(x)}
=
\Bigl( \arg\min_{\gamma \in \Pi_{\mu,\nu}} C(\gamma) \Bigr) \cap \mathrm{Perm}(N),
\end{equation}
where $\mathrm{Perm}(N)$ denotes the set of $N\times N$ permutation matrices. Hence the (MK) problem, which is a relaxation of (M), admits as a special case the Monge solutions when permutation-type optimal plans exist.
\end{rem}

\subsubsection{Wasserstein Metric}

\begin{definition}
We denote by $P(\R^n)$ the set of (Borel) probability measures on $\R^n$. We define the set of atomic probability measures by
\begin{equation}
A(\R^n) := \left\{ \mu \in P(\R^n) \,\middle|\, \exists N \in \N^*,\, (x_i)_{i=1}^N \in (\R^n)^N,\, (\mu_i)_{i=1}^N \in [0,1]^N,\, \sum_{i=1}^N \mu_i=1,\; \mu = \sum_{i=1}^N \mu_i \,\delta_{x_i} \right\}.
\end{equation}
\end{definition}

\begin{definition}
Let $p \geq 1$. For finite subsets $X, Y \subset \R^n$, and $\mu, \nu$ given in the (MK) setting as
\[
\mu = \sum_{x \in X} \mu_x \delta_x, \qquad \nu = \sum_{y \in Y} \nu_y \delta_y,
\]
we set $c := (\| y - x \|^p)_{(x,y) \in X \times Y}$ and define $W_p : A(\R^n) \times A(\R^n) \to \R_+$ by
\begin{equation}
W_p(\mu,\nu) := \left( \min_{\gamma \in \Pi_{\mu,\nu}} C(\gamma) \right)^{1/p}
= \left( \min_{\gamma \in \Pi_{\mu,\nu}} \sum_{(x,y) \in X \times Y} \gamma_{x,y} \, \|y - x\|^p \right)^{1/p}.
\end{equation}
\end{definition}

\begin{rem}
The quantity $W_p$ defines a metric on $A(\R^n)$ and can be extended to more general classes of probability measures. It captures an intuitive notion of the ``minimal total displacement'' required to transform one mass distribution into another.
\end{rem}

\subsection{Generalized Entropy}

The notion of (Shannon) entropy \cite{shannon1948mathematical} is fundamental in information theory and statistics: it quantifies the dispersion of a distribution and the expected information content of an observation. For a discrete probability vector $p=(p_i)$, this entropy is given by $H(p) = -\sum_i p_i\log p_i$. In the continuous case, one can define an analogous \emph{differential entropy}, although its interpretation is more subtle due to its dependence on the underlying reference measure and units.

To apply entropic regularization in optimal transport, it is convenient to introduce a divergence that compares a probability measure to a reference one. This leads to the Kullback--Leibler divergence \cite{ali1966general}.

\begin{definition}
Let $M$ be a compact subset of $\R^d$. Let $P$ and $Q$ be two probability measures on $M$ that are absolutely continuous with respect to a common reference measure (e.g.\ Lebesgue), with corresponding densities $p$ and $q$. The Kullback--Leibler divergence of $P$ with respect to $Q$ is defined as
\begin{equation}
\mathrm{KL}(P\|Q) = \int_M p(x)\,\log\!\left(\frac{p(x)}{q(x)}\right)\,dx,
\end{equation}
with the usual conventions $0\log 0 = 0$ and $p(x)\log(p(x)/q(x))=+\infty$ if $p(x)>0$ and $q(x)=0$.
\end{definition}

In the discrete case $P=(p_i)_{i}$ and $Q=(q_i)_i$, the definition becomes
\begin{equation}
\mathrm{KL}(P\|Q) = \sum_i p_i\,\log\!\left(\frac{p_i}{q_i}\right).
\end{equation}

\subsection{Sinkhorn Algorithm}

As seen in subsection \ref{transopt}, minimizing the Wasserstein distance is not an easy problem : the discrete optimal transport problem in the (MK) formulation is a finite-dimensional linear program and is therefore convex, but generic solvers can have a high computational cost when the number of points grows. 
A popular regularized variant replaces the original cost $C(\gamma)$ by a cost plus entropic penalty and leads to a more numerically tractable problem. In its simplest symmetric form, one considers
\begin{equation}
\min_{\gamma \in \Pi_{\mu,\nu}} \sum_{(x,y)} c_{x,y}\,\gamma_{x,y} + \varepsilon\,\mathrm{KL}(\gamma\|\mu\otimes\nu),
\end{equation}
where $\varepsilon>0$ is a regularization parameter and $\mu\otimes\nu$ is the product measure. The associated Gibbs kernel
\begin{equation}
K_{x,y} = \exp\!\left(-\frac{c_{x,y}}{\varepsilon}\right)
\end{equation}
plays a central role in the resulting algorithm.

The Sinkhorn algorithm (also known as the iterative proportional fitting procedure) can be written in matrix form as follows. One seeks a coupling $\pi$ of the form
\begin{equation}
\pi = \mathrm{diag}(u)\,K\,\mathrm{diag}(v),
\end{equation}
where $u$ and $v$ are scaling vectors chosen so that the marginals of $\pi$ match $\mu$ and $\nu$:
\[
\pi \mathbf{1} = \mu,\qquad \pi^\top \mathbf{1} = \nu.
\]
Starting from an initial guess (for instance $v^{(0)} = \mathbf{1}$), the classical Sinkhorn iterations update $u$ and $v$ by alternating normalizations:
\begin{align}
u^{(\ell+1)} &= \mu \oslash (K v^{(\ell)}),\\
v^{(\ell+1)} &= \nu \oslash (K^\top u^{(\ell+1)}),
\end{align}
where $\oslash$ denotes entrywise division. One can monitor convergence via the marginal residuals
\[
\|\pi^{(\ell)}\mathbf{1} - \mu\|_1, \qquad \|(\pi^{(\ell)})^\top \mathbf{1} - \nu\|_1,
\]
with $\pi^{(\ell)} = \mathrm{diag}(u^{(\ell)}) K \mathrm{diag}(v^{(\ell)})$.

Each iteration involves a small number of matrix--vector products with the kernel $K$. For a dense $n\times n$ kernel, the cost of one iteration is $O(n^2)$. The total complexity depends on the required accuracy and on $\varepsilon$; recent analyses give bounds of the form $O(n^2 \log n / \varepsilon^2)$ in various regimes~\cite{cuturi2013sinkhorn,NTsink}, which is typically much lower than that of solving the unregularized transport problem by generic linear programming methods.

\subsection{The Heat Equation}

The heat equation is a fundamental partial differential equation. In three spatial dimensions it reads
\begin{equation}
\frac{\partial f_t}{\partial t} = \frac{\partial^2 f_t}{\partial x_1^2} + \frac{\partial^2 f_t}{\partial x_2^2} + \frac{\partial^2 f_t}{\partial x_3^2} = \Delta f_t,
\end{equation}
with a prescribed initial condition $f_0$.

On a finite graph or mesh, one typically replaces the continuous Laplacian $\Delta$ by a discrete Laplacian operator $L$ acting on functions defined on vertices. In matrix notation, for a vector-valued signal $X_t$ (e.g.\ stacking 3D coordinates), a standard linear diffusion equation takes the form
\begin{equation}
\label{eq:discrete-heat}
\frac{\partial X_t}{\partial t} = - L X_t,
\end{equation}
where the sign is chosen so that the flow is smoothing (the Laplacian is positive semi-definite and $-L$ generates a contraction semigroup).

\subsection{Cholesky Factorization}

The Cholesky factorization used here is a linear algebra technique that simplifies the solution of symmetric positive definite systems.

\begin{prop}
Let $A$ be a real, symmetric, positive definite matrix. Then there exists a unique lower triangular matrix $L$ with strictly positive diagonal entries such that
\begin{equation}
A = L\,L^\top.
\end{equation}
\end{prop}

Solving linear systems $A x = b$ then reduces to solving two triangular systems with $L$ and $L^\top$, which is more efficient and numerically stable than inverting $A$ explicitly.

\subsection{Denoising Meshes and Images Using the Sinkhorn Algorithm}

\begin{figure}[h]
      \centering
      \includegraphics[scale=0.3]{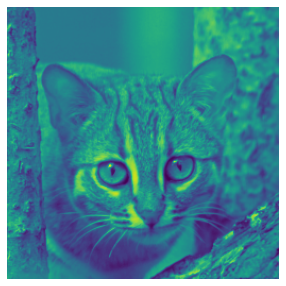}
      \includegraphics[scale=0.3]{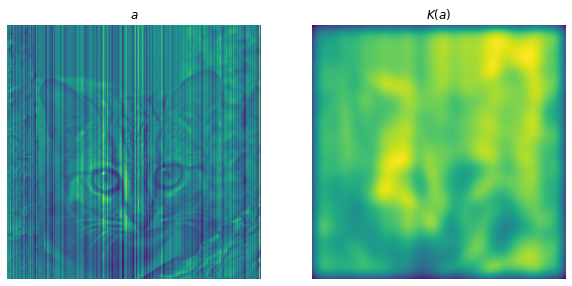}
      \includegraphics[scale=0.3]{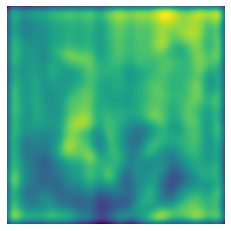}
      \caption{From left to right: a photo of a cat; the same photo corrupted with white noise of amplitude $\rho = 0.2$; the kernel of the noisy image; the barycenter $b$ sought after one iteration of the algorithm.}
\end{figure}

It is possible in principle to compute a barycenter (or Wasserstein mean) of noisy images or meshes in order to reduce noise. In a discrete setting, this amounts to solving
\begin{equation}
\min_b \sum_{k=1}^R W_\gamma(a_k, b),
\end{equation}
where $(a_k)_{k=1}^R$ are input distributions and $W_\gamma$ is a regularized optimal transport cost. This problem can be addressed with variants of the Sinkhorn algorithm~\cite{benamou2015iterative,SOL2015}.

In practice, the choice of kernel $K$ and regularization parameter $\varepsilon$ has a strong impact on the result. In a preliminary image denoising experiment with a simple Gibbs kernel, the visual quality was not very satisfactory, which suggests that more adapted kernels or multiscale strategies are needed for this approach to compete with simpler methods.

\section{Three Mesh Denoising Methods}

We now present the three mesh denoising methods that we test in detail. For this, we introduce the (normalized) adjacency matrix of a mesh.

\begin{definition}[Adjacency and related matrices]
Let $\mathcal{E}$ be the set of edges of $\mathcal{M}$. We define the (possibly directed) adjacency matrix $W$ by
\begin{equation}
W_{i,j} =
\begin{cases}
1 & \text{if } (i,j) \in \mathcal{E}, \\
0 & \text{otherwise.}
\end{cases}
\end{equation}
We define the degree vector $d \in \N^n$ by
\begin{equation}
d_i = \sum_j W_{i,j},
\end{equation}
and the diagonal degree matrix $D = \mathrm{diag}_i(d_i)$. The (row-)normalized adjacency matrix is then defined by
\begin{equation}
\tilde{W} = D^{-1} W.
\end{equation}
We also define the (unnormalized) graph Laplacian
\begin{equation}
L := D - W,
\end{equation}
and the normalized Laplacian
\begin{equation}
\tilde{L} := D^{-1} L = I_n - \tilde{W}.
\end{equation}
\end{definition}

\begin{rem}
When $\tilde{W}$ is row-stochastic, left-multiplication by $\tilde{W}$ averages each vertex with its neighbors and can be interpreted as a simple low-pass filtering step. Conversely, the Laplacian $L$ emphasizes local differences and is often associated with high-frequency components. These interpretations become precise in a spectral graph framework.
\end{rem}

\subsection{Filtering}

Let $X \in \R^{n\times 3}$ denote the matrix collecting the 3D coordinates of the noisy vertices. A simple filtering operation consists in repeatedly applying a neighborhood averaging operator, for instance by
\begin{equation}
X^{(\ell+1)} = \tilde{W} X^{(\ell)},
\end{equation}
starting from $X^{(0)} = X$. In our implementation, we use an equivalent formulation that multiplies $X$ on the right by $\tilde{W}^\top$, which corresponds to a transpose convention.

At each iteration, this requires multiplying an $n\times n$ matrix by an $n\times 3$ matrix. If one ignores sparsity, the cost is $O(n^2)$ operations per iteration (the factor $3$ being constant). On the actual meshes we consider, the adjacency is sparse with bounded degree, so the true cost per iteration is closer to $O(m)$ operations, where $m$ is the number of edges.

\subsection{Linear Heat Diffusion}

We use the normalized Laplacian $\tilde{L} = I_n - \tilde{W}$ as above. The idea is to approximate a denoised mesh by the solution of a discrete heat equation of the form~\eqref{eq:discrete-heat}. Using an explicit Euler scheme with time step $\tau>0$ leads to the recurrence
\begin{equation}
X^{(\ell+1)} = X^{(\ell)} - \tau \tilde{L} X^{(\ell)} = (1-\tau)\,X^{(\ell)} + \tau \tilde{W} X^{(\ell)}.
\end{equation}
This is a convex combination of the identity and the averaging operator, and for sufficiently small $\tau$ it defines a stable diffusion process.

For $\tau = 1$, this iteration reduces to a pure filtering step of the type described above. For intermediate values of $\tau$, especially on smaller meshes, we observe that an appropriate choice of $\tau$ can improve the denoising quality for a given number of iterations.

Each iteration again involves one multiplication by $\tilde{W}$ (and a few vector operations). Treating the operators as dense matrices and the signal as $n\times 3$, the cost is $O(n^2)$ per iteration, but with sparse adjacency the practical cost scales with the number of edges $m$.

\subsection{Sobolev Regularization}

A different approach is to define the denoised mesh as the minimizer of a penalized least-squares functional of the form
\begin{equation}
\min_{Y \in \R^{n\times 3}} \|Y - X\|_F^2 + \mu\,\|L^{1/2}Y\|_F^2,
\end{equation}
where $\mu>0$ is a regularization parameter and $\|\cdot\|_F$ denotes the Frobenius norm. The Euler--Lagrange equation leads to a linear system
\begin{equation}
(I_n + \mu L)\,Y = X.
\end{equation}
Equivalently, in transposed notation,
\begin{equation}
X_\mu^\top = (I_n + \mu L)^{-1} X^\top,
\end{equation}
where $X_\mu$ denotes the denoised vertex matrix.

Since $L$ is symmetric positive semi-definite, the matrix $I_n + \mu L$ is symmetric positive definite for all $\mu>0$, so it admits a Cholesky factorization $(I_n + \mu L) = LL^\top$. One can thus solve the system efficiently by forward/backward substitution for each coordinate column of $X$.

If $(I_n + \mu L)$ is treated as a dense matrix, the Cholesky factorization costs $O(n^3)$ operations, and applying the factorization to the $3$ coordinate columns costs $O(n^2)$ per column. On sparse meshes with bounded degree, the effective complexity is significantly reduced by exploiting sparsity and suitable orderings, and in our experiments the cost of this one-shot solve for each mesh is much smaller than that of running many iterations of the iterative methods. Importantly, once the factorization has been performed for a given $\mu$, changing $X$ does not require recomputing it.

The higher $\mu$ is, the stronger the smoothing effect. In our experiments, we tune $\mu$ to optimize the SNR on each dataset.

\section{Empirical Comparison of These Methods on Two Different Meshes}

To empirically compare the three methods, we apply them to two different meshes representing an elephant (24,955 vertices, 49,918 faces, and 74,877 edges) and the bust of Nefertiti (299 vertices, 562 faces, and 860 edges) \cite{NTmeshd,NTmeshp}.

\begin{figure}[h]
      \centering
      \includegraphics[scale=0.3]{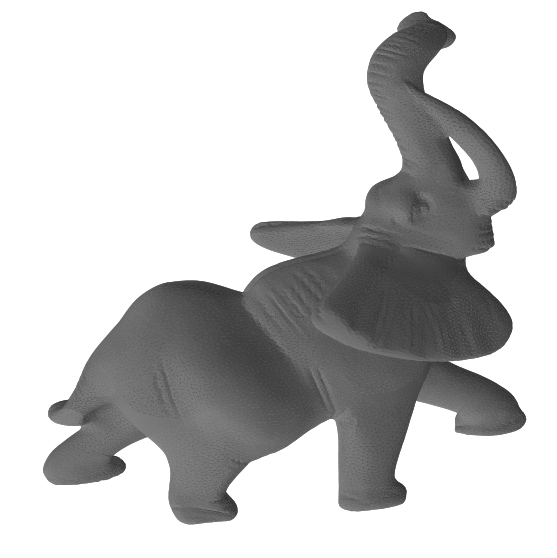}
      \includegraphics[scale=0.3]{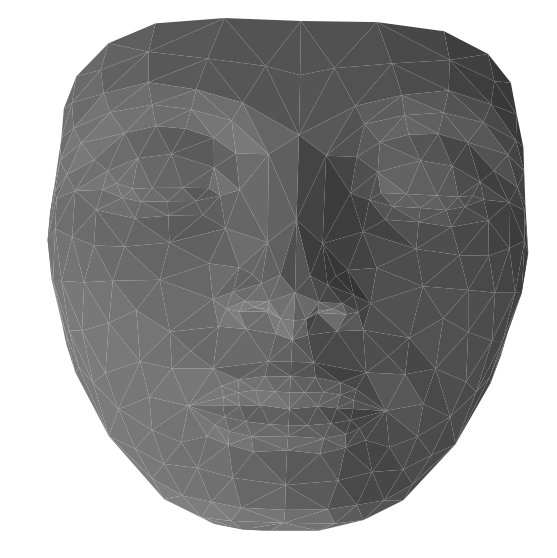}
      \caption{The original meshes, an elephant \cite{NTmeshd} and the bust of Nefertiti \cite{NTmeshp}.}
\end{figure}

We add white Gaussian noise to these meshes by displacing each vertex along its normal direction:
\begin{equation}
x_i = x_{0,i} + \rho \,\epsilon_i\, N_i \in \R^3,
\end{equation}
with $\epsilon_i \sim \mathcal{N}(0,1)$, $N_i \in \R^3$ the unit normal at vertex $i$, and $\rho > 0$ a parameter controlling the noise level. We compare the methods in terms of effectiveness and empirical execution time. To measure effectiveness, we calculate the error in decibels (dB) using the signal-to-noise ratio (SNR):
\begin{equation}
\mathrm{SNR}(X,Y) := -20 \log_{10}\!\left( \frac{\| X - Y \|_F}{\| Y \|_F} \right),
\end{equation}
where $X$ and $Y$ are the vertex matrices of the original and denoised meshes, respectively. A higher SNR indicates a denoised mesh closer to the original (with $\mathrm{SNR} = +\infty$ corresponding to a perfect reconstruction).

\subsection{Comparative Elements and Interpretations}

All three denoising methods are able to transform severely corrupted meshes into shapes that are visually recognizable to a human observer when displayed on a screen. Quantitatively, we observe systematic differences between the two meshes, which we attribute mainly to their very different sizes, as both meshes have similar connectivity characteristics in relative terms. The three methods perform substantially better on the large mesh. As expected, the higher the noise level, the harder it is to recover the original mesh.

Filtering and linear heat diffusion yield very similar SNRs for comparable parameter choices. Sobolev regularization produces slightly lower SNR values (typically by $1$--$2$ dB) but is significantly faster in our experiments, because it solves a single linear system for each mesh instead of performing many iterations.

\begin{figure}
\label{prosob}
      \centering
      \includegraphics[scale=0.38]{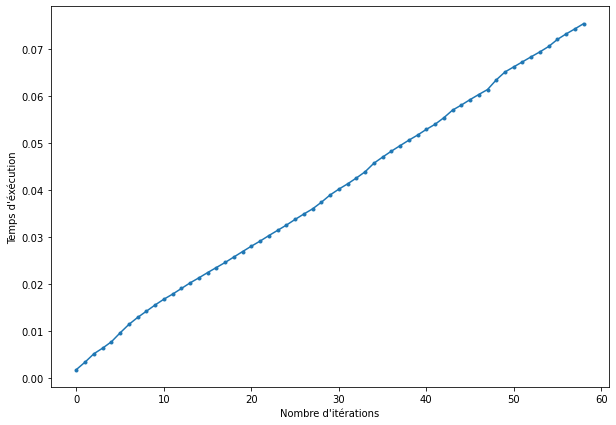}
      \includegraphics[scale=0.38]{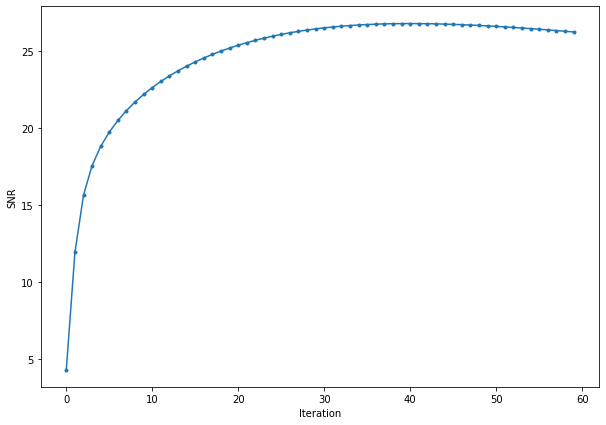} \\
      \includegraphics[scale=0.37]{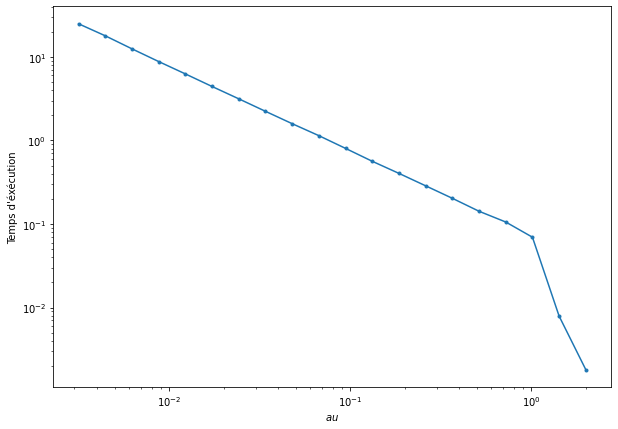}
      \includegraphics[scale=0.37]{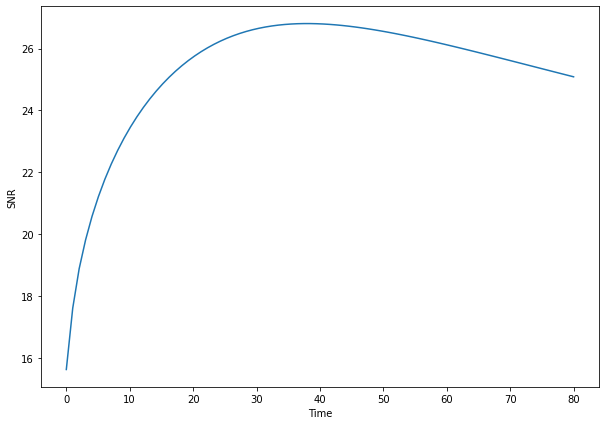}
      \caption{\textbf{Top:} Processing time versus the number of filtering iterations (left, in seconds) and denoising level versus the number of iterations (right), for the elephant mesh with noise $\rho = 0.2$. \\ \textbf{Bottom:} Processing time versus the choice of parameter $\tau$ to achieve the best possible result (left, in seconds) and denoising level versus the number of iterations for the most effective and fastest algorithm, with $\tau = 1.01$, maximum achieved at $T = 37.4$, with an execution time of 70 ms (right), for the elephant mesh with noise $\rho = 0.2$. \\ Note that to achieve a denoising level similar to Sobolev regularization (25 dB), it takes 20 ms for filtering, as well as for the heat equation method. Since the execution time of the Sobolev algorithm does not depend on the parameter $\mu$, we have all the necessary information for comparison.}
\end{figure}

\begin{table}[h]
\label{datamaillage}
\centering
\begin{tabular}{c||c|c|c|c}
Mesh & Noisy Image & Filtering & Heat & Sobolev \tabularnewline
\midrule 
Elephant, $\rho = 0.015$ & 26.77 & 39.82 & 39.81 & 38.13 \tabularnewline
Elephant, $\rho = 0.2$ & 4.34 & 26.81 & 26.81 & 25.27 \tabularnewline
\midrule 
Nefertiti, $\rho = 0.015$ & 41.76 & 41.76 & 41.83 & 21.77 \tabularnewline
Nefertiti, $\rho = 0.2$ & 19.33 & 23.17 & 24.17 & 22.95 \tabularnewline
Nefertiti, $\rho = 1$ & 5.15 & 14.56 & 14.62 & 13.86 \tabularnewline
\end{tabular} 
\caption{Best $SNR$ (in dB) obtained by the different denoising methods.}
\end{table}

\begin{table}[h]
\centering
\begin{tabular}{c||c|c|c}
Mesh & Filtering & Heat & Sobolev \tabularnewline
\midrule 
Elephant, $\rho = 0.015$ & 14.2 $\pm$ 0.449 & 9.66 $\pm$ 0.098 & 0.0355 $\pm$ 0.002 \tabularnewline
Elephant, $\rho = 0.2$ & 50.8 $\pm$ 1.38 & 70.7 $\pm$ 0.6 & 0.0349 $\pm$ 0.0046 \tabularnewline
\midrule 
Nefertiti, $\rho = 0.015$ & 0 & 0.106 $\pm$ 0.0134 & 0.0831 $\pm$ 0.0178 \tabularnewline
Nefertiti, $\rho = 0.2$ & 0.0447 $\pm$ 0.0022 & 0.507 $\pm$ 0.051 & 0.101 $\pm$ 0.0275 \tabularnewline
Nefertiti, $\rho = 1$ & 0.303 $\pm$ 0.0351 & 0.465 $\pm$ 0.048 & 0.0787 $\pm$ 0.016 \tabularnewline
\end{tabular} 
\caption{Machine usage time (in milliseconds, ms) for the best result obtained by the different denoising methods, calculated using the \textit{timeit} method.}
\end{table}

Sobolev regularization is clearly the fastest method in our experiments, often by more than an order of magnitude in wall-clock time. Moreover, by examining the curves showing execution time versus the number of iterations (Figure~\ref{prosob}), we observe that for comparable SNR levels Sobolev regularization is systematically faster than the best filtering or heat diffusion configurations. On the elephant mesh, the value of $\tau$ that maximizes SNR for the heat equation method is frequently close to $1$, which makes this scheme behave similarly to repeated filtering while incurring a higher computational overhead.

\begin{figure}[h]
\label{bruittau}
      \centering
      \includegraphics[scale=0.2]{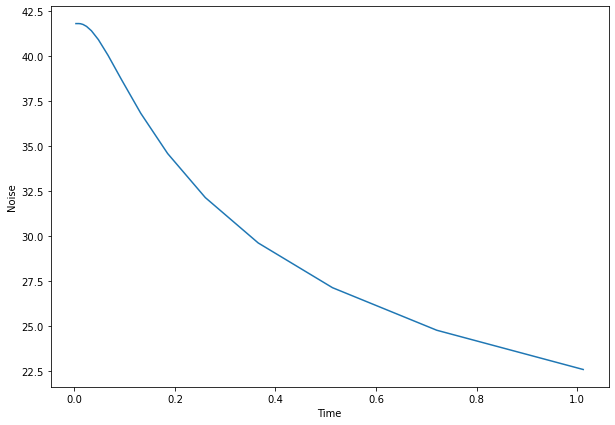}
      \includegraphics[scale=0.2]{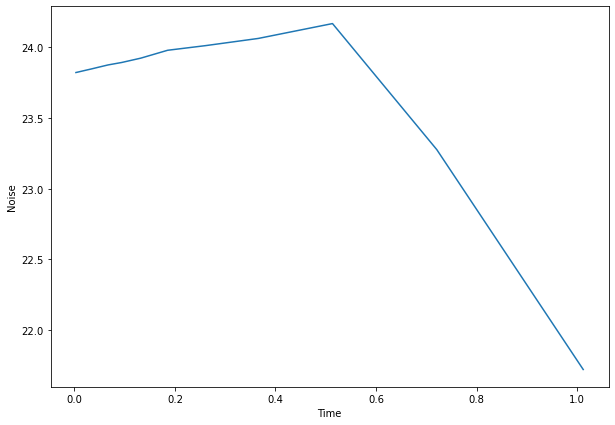}
      \includegraphics[scale=0.2]{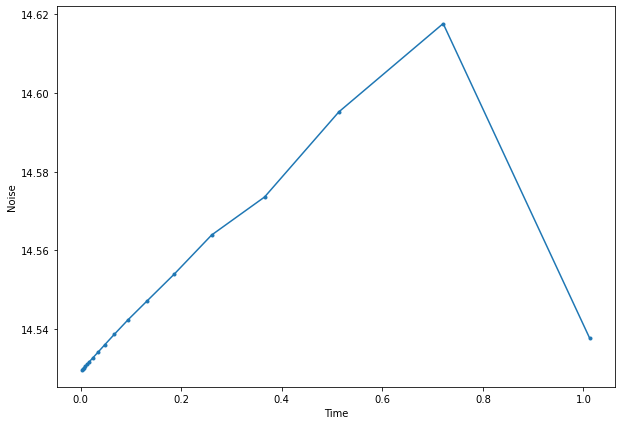}
      \caption{Best noise reduction after processing by heat diffusion on the Nefertiti busts. From left to right: $\rho = 0.015$, $\rho = 0.2$, $\rho = 1$.}
\end{figure}

On the Nefertiti bust, however, the heat diffusion method can achieve a slightly better SNR for an intermediate value of $\tau$ (around $0.5$ for $\rho = 0.2$ and around $0.7$ for $\rho = 1$), see Figure~\ref{bruittau}. This suggests that on small meshes there can be more sensitivity to parameter choices.

We also note a difference in optimal parameters between the two meshes for the same noise level. Generally, for the large mesh, the optimal regularization parameters are higher: for the Nefertiti mesh with an initial SNR of $5.15$ dB, values such as $\mu = 2$ for Sobolev regularization, $3$ filter passes, and a diffusion time of order $3$ are sufficient to reach the empirical maximum. As the noise level increases, more iterations or stronger regularization are required. For the elephant mesh, with an initial SNR of $4.34$ dB at $\rho = 0.2$, the best denoising we observed is obtained around $\mu = 51$, $40$ iterations of filtering, and a diffusion time of about $40$ units for the heat equation method.

\section{Conclusion}

Our study shows that, among the three mesh-based methods studied in detail, Sobolev regularization offers the best trade-off between accuracy and speed: it achieves SNR values close to those of the best filtering and heat diffusion configurations while requiring much less computation time. Filtering and heat diffusion achieve very similar SNRs, with heat diffusion sometimes offering marginal improvements on the small mesh when its parameters are carefully tuned.

\begin{figure}
      \centering
      \includegraphics[scale=0.2]{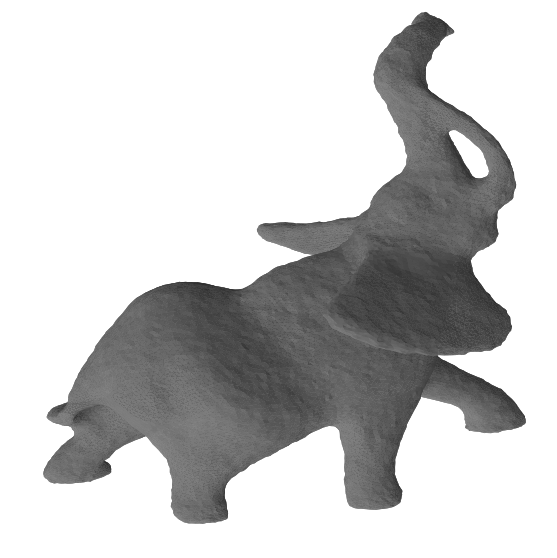}
      \includegraphics[scale=0.2]{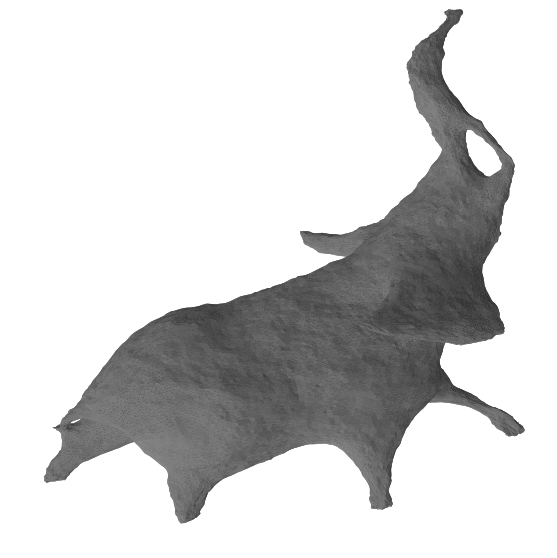} \\
      \includegraphics[scale=0.2]{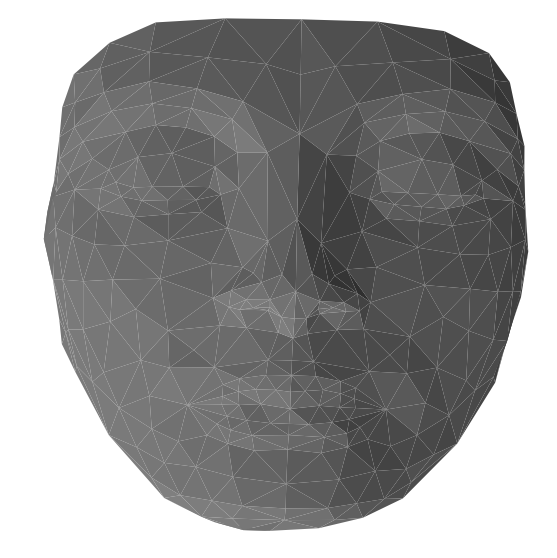}
      \includegraphics[scale=0.2]{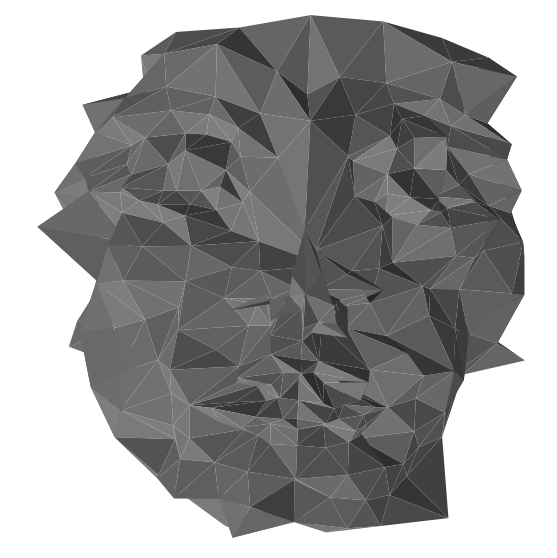}
      \includegraphics[scale=0.2]{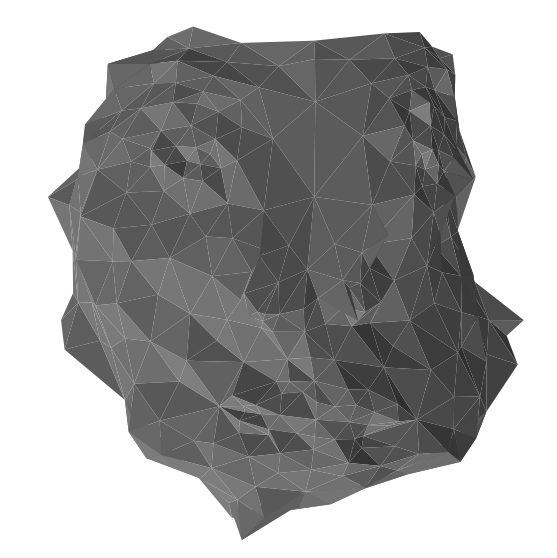}
      \caption{The various meshes studied, denoised using Sobolev regularization. From left to right: $\rho = 0.015$, $\rho = 0.2$, $\rho = 1$.}
\end{figure}

Three main limitations emerge from our experiments. First, all three methods are markedly less effective on the small Nefertiti mesh than on the large elephant mesh, which suggests that mesh resolution strongly constrains the maximum achievable SNR. Second, the maximum SNR appears bounded away from perfect reconstruction: for instance, the elephant's eye is not recovered in any of the denoised meshes, so the reconstruction remains perceptibly imperfect. Third, for some low-noise configurations, additional processing brings limited SNR gains and can occasionally slightly degrade the quantitative measure, which calls for more adaptive or data-driven strategies.

One interesting avenue to improve mesh denoising would be to use the Sinkhorn algorithm for barycentric interpolation of geometric data, as proposed in~\cite{SOL2015}. The main challenges would be the design of kernels adapted to mesh geometry (for example based on discrete heat kernels) and the adaptation of the framework to signals supported on irregular meshes with varying resolution.

\paragraph{Acknowledgments} Thanks to Gabriel Peyré for the valuable resources used here.

\paragraph{Competing Interests} I declare no competing interests.

\paragraph{Supplementary Information} The code is available at the following address: \href{https://github.com/cvt8/Mesh-denoising}{https://github.com/cvt8/Mesh-denoising}.

\addcontentsline{toc}{section}{References}
\printbibliography

@article{benamou2015iterative,
  title={Iterative Bregman projections for regularized transportation problems},
  author={Benamou, Jean-David and Carlier, Guillaume and Cuturi, Marco and Nenna, Luca and Peyr{\'e}, Gabriel},
  journal={SIAM Journal on Scientific Computing},
  volume={37},
  number={2},
  pages={A1111--A1138},
  year={2015},
  publisher={SIAM}
}

@article{cuturi2013sinkhorn,
  title={Sinkhorn distances: Lightspeed computation of optimal transport},
  author={Cuturi, Marco},
  journal={Advances in neural information processing systems},
  volume={26},
  pages={2292--2300},
  year={2013}
}

@article{SOL2015,
  title={Convolutional wasserstein distances: Efficient optimal transportation on geometric domains},
  author={Solomon, Justin and De Goes, Fernando and Peyr{\'e}, Gabriel and Cuturi, Marco and Butscher, Adrian and Nguyen, Andy and Du, Tao and Guibas, Leonidas},
  journal={ACM Transactions on Graphics (TOG)},
  volume={34},
  number={4},
  pages={1--11},
  year={2015},
  publisher={ACM New York, NY, USA}
}

@online{NTsink,
author = {Peyré, Gabriel},
title = {Numerical tour : Entropic regularisation of optimal transport},
year = {2019},
url = {http://www.numerical-tours.com/python/},
}

@online{NTmeshd,
author = {Peyré, Gabriel},
title = {Numerical tour : Mesh denoising},
year = {2019},
url = {http://www.numerical-tours.com/python/},
}

@online{NTmeshp,
author = {Peyré, Gabriel},
title = {Numerical tour : Mesh parametrisation},
year = {2019},
url = {http://www.numerical-tours.com/python/},
}

@book{OTAM,
	author = {Filippo Santambrogio},
	title = {Optimal Transport for Applied Mathematicians – Calculus of Variations, PDEs and Modeling},
	date = {2015},
}

@article{shannon1948mathematical,
  title={A mathematical theory of communication},
  author={Shannon, Claude Elwood},
  journal={The Bell system technical journal},
  volume={27},
  number={3},
  pages={379--423},
  year={1948},
  publisher={Nokia Bell Labs}
}

@article{ali1966general,
  title={A general class of coefficients of divergence of one distribution from another},
  author={Ali, Syed Mumtaz and Silvey, Samuel D},
  journal={Journal of the Royal Statistical Society: Series B (Methodological)},
  volume={28},
  number={1},
  pages={131--142},
  year={1966},
  publisher={Wiley Online Library}
}

\end{document}